\documentclass[%
 aip,
 jmp %
 amsmath,amssymb,
 reprint,%
]{revtex4-1}

\usepackage{graphicx}
\usepackage{dcolumn}
\usepackage{bm}

\begin{document}

\preprint{AIP/123-QED}

\title{High-Temperature Superconducting Multi-Band Radio-Frequency Metamaterial Atoms} 

\author{Behnood G. Ghamsari}
\affiliation{\footnotesize Center for Nanophysics and Advanced
Materials, Department of Physics, University of Maryland, College
Park, MD, 20742-4111, USA}
\author{John Abrahams}
\affiliation{\footnotesize Center for Nanophysics and Advanced
Materials, Department of Physics, University of Maryland, College
Park, MD, 20742-4111, USA}
\author{Steven M. Anlage}
\affiliation{\footnotesize Center for Nanophysics and Advanced
Materials, Department of Physics, University of Maryland, College
Park, MD, 20742-4111, USA}

\date{\today}

\begin{abstract}
We report development and measurement of a micro-fabricated compact
high-temperature superconducting (HTS) metamaterial atom operating at a frequency as low as $\sim$ 53MHz.
The device is a planar spiral resonator patterned out of a {YBa$_2$Cu$_3$O$_{7-\delta}$} (YBCO) thin film
with the characteristic dimension of $\sim \lambda_0/1000$, where $\lambda_0$ is the free-space wavelength of the fundamental resonance.
While deployment of an HTS material enables higher operating temperatures and greater tunability, it has not compromised the quality of our spiral metamaterial atom and
a Q as high as $\sim 1000$ for the fundamental mode, and $\sim 30000$ for higher order modes, are achieved up to 70K.
Moreover, we have experimentally studied the effect of the substrate by comparing the performance of similar devices on different substrates.
\end{abstract}


\maketitle

Realization of metamaterials based on sub-wavelength artificial electromagnetic structures has attracted much attention and efforts for
a variety of applications\cite{Shalaev,Smith}.
This interest applies virtually to the entire span of the electromagnetic spectrum from radio-frequency (RF) up to visible and ultraviolet wavelengths.
Nevertheless, at the low-frequency limit, i.e. in the RF regime, where the free-space wavelength of signals ranges from tens of centimeters to meters, the constituent elements of metamaterials, referred to as the metamaterial atoms, are often bulky, which hinders the implementation of scalable RF metamaterials.
In addition, RF resonant structures traditionally exhibit low quality factors due to significant dissipation in their bulky structures\cite{Wiltshire,Wiltshire2}.

Since both requirements of scalability/compactness and low dissipation have proved to be challenging to achieve for RF metamaterials, few studies have concerned metamaterials at the low-frequency part of the electromagnetic spectrum\cite{Chen}. Nonetheless, RF metamaterials have been sought to improve the performance of magnetic resonance imaging (MRI) devices\cite{Wiltshire3,Freire}, magnetoinductive lenses\cite{Freire2}, microwave antennas\cite{Ziolkowski}, delay-lines\cite{Freire3}, and resonators\cite{Engheta}.

Insofar as obtaining RF metamaterials substantially relies on the development of compact and scalable metamaterial atoms that are amenable to conventional micro-fabrication techniques, many superconducting structures have been recently introduced and tested for metamaterial applications in the RF/microwave regime\cite{Ricci1, Ricci2, Ricci3, Salehi, Fedotov, Chen2, Wang, Anlage,Tsiatmas,Savinov}, motivated by their low-losses and deep sub-wavelength sizes.

Recently, we have demonstrated superconducting RF metamaterials based on Niobium (Nb) spiral resonators as a viable means to realize efficient metamaterials at low frequencies presenting both a compact physical structure and low loss\cite{Kurter1,Kurter2}. Given that Nb is a low-temperature superconductor (LTS) whose transition temperature is $\sim $9.2K, development of analogous superconducting metamaterial atoms capable of functioning at the temperature of liquid nitrogen, 77K, is clearly a technological advantage.
Furthermore, accessing a wider range of superconducting temperature allows greater convenience and effectiveness in tuning the properties of the metamaterial atom by means of controlling the operating temperature.

The same thermal mechanism may be employed to dynamically control the properties of the device through optical illumination, as has been recently demonstrated in optically tunable superconducting microwave resonators and delay-lines\cite{Atikian}. Thus, replacing Nb, which possess a highly-reflective surface, with an HTS material will also enhance optical coupling and thereby will improve dynamic tunability of the metamaterial device by means of an optical control signal.

Our YBCO spiral resonator, while maintaining a high quality factor, Q, functions at the deep sub-wavelength regime where its physical dimension is 1000 times smaller than the free space wavelength of the fundamental mode, $\lambda_0$.
These characteristics are comparable with natural atomic resonators such as a Hydrogen atom, which has a similar size to wavelength ratio for visible light and possesses a Q around $15\times10^3$ for the doppler broadened $H_\alpha$ line in the sun.
Such high quality factors and deep sub-wavelength operation can only be attained through compact superconducting resonators. More compact superconducting resonators at even deeper sub-wavelength scales may be obtained by nano-patterning the spiral resonator in order to further enhance the kinetic inductance of the device and lower its resonance frequency.

\begin{figure}
\centering{\includegraphics[width=3.5in]{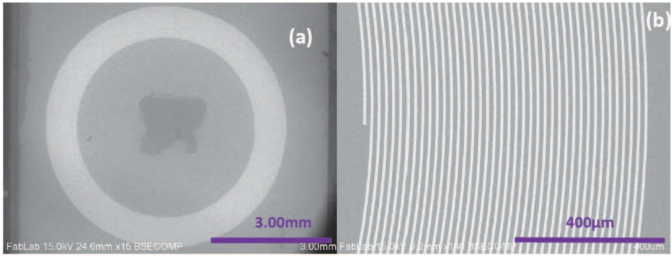}
} \caption{SEM images of a 6mm-diameter YBCO spiral resonator with 40 turns of 10$\mu m$-wide lines and 10$\mu m$ spacing.} \label{Spiral}
\end{figure}

Our HTS metamaterial atom comprises a 6$mm$-diameter planar spiral resonator consisting of 40 turns of nominally 10$\mu m$-wide lines separated by 10$\mu m$ gaps patterned out of a 300$nm$-thick YBCO film on a sapphire substrate, as depicted in figure \ref{Spiral}. Photolithography has been performed by spin-coating the sample by Hexamethyldisiloxane (HMDS) at 4000rpm for 60 seconds followed by spinning the S1813 photoresist at 4000rpm for 60 seconds and soft baking at 100$^\circ$C for 1 minute. The coated sample is then exposed to 405nm light at 8$mW/cm^2$ for 8 seconds, developed with un-diluted CD-30 developer for 20 seconds, and hard baked at 120$^\circ$C for 5 minutes. The sample is ultimately flood etched in 2\% diluted phosphoric acid for 1 minute.

\begin{figure}
\centering{\includegraphics[width=3.5in]{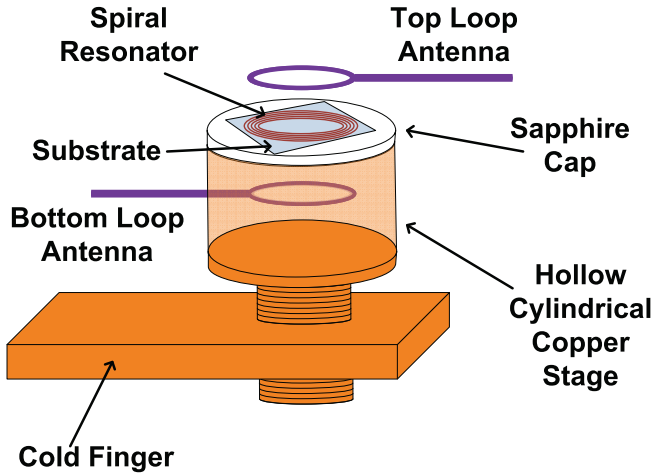}
} \caption{Schematic of the measurement setup. The loop antennas are connected to an RF network analyzer at room temperature, while the cold finger is connected to a Gifford-McMahan cryocooler.} \label{Setup}
\end{figure}

Figure \ref{Setup} illustrates the measurement setup. The spiral resonator rests on a 2.54cm diameter sapphire disk which caps a hollow cylindrical copper holder providing the thermal link between the sample and the cold finger of a DE-210SF Advanced Research System closed-cycle optical cryostat. RF-microwave characteristics of the resonator are measured by means of two parallel loop antennas. The first antenna approaches the device directly from the top and the other loop antenna is brought in from the bottom through a slit in the body of the cylindrical copper holder. The loop antennas are formed by connecting the inner and outer conductors of a semi-rigid coaxial cable. The scattering parameters (S-parameters) are measured by an Agilent E8364C network analyzer.

\begin{table*}
\caption{\label{fQ}The experimental resonance characteristics of the low-order modes below 1GHz for the YBCO spiral resonator at 10K with different substrates.}
\begin{ruledtabular}
\begin{tabular}{ccccccc}
 &\multicolumn{2}{c}{$MgO$}&\multicolumn{2}{c}{$Al_2O_3$}&\multicolumn{2}{c}{$LaAlO_3$}\\
 $n$&$f_n (MHz)$&$Q_n$&$f_n (MHz)$&$Q_n$&$f_n (MHz)$&$Q_n$
\\ \hline
    1&54.9&650&53.4&1040&77.7&380\\
    2&157.6&9000&154.7&20900&210.5&2630\\
    3&255.1&3000&250.2&5200&265.1&1190\\
    4&358.4&14500&350.8&3200&376.0&17700\\
    5&458.2&7800&447.6&10300&453.0&730\\
    6&560.8&30000&546.7&28900&643.2&6500\\
    7&661.2&11500&643.4&13700&746.1&950\\
    8&763.4&27700&741.5&21500&932.6&11400\\
    9&863.9&16500&837.6&16200&&\\
    10&965.9&17800&934.9&9300&&\\
\end{tabular}
\end{ruledtabular}
\end{table*}

\begin{figure}
\centering{\includegraphics[width=3.5in]{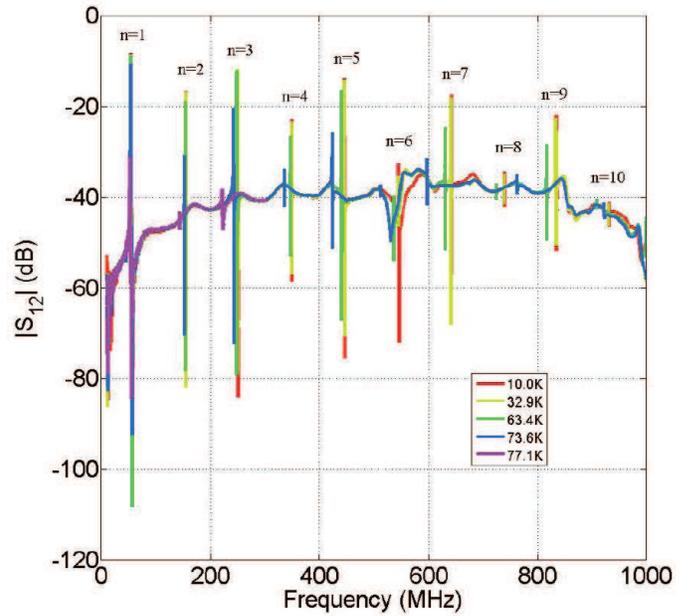}
} \caption{Transmission spectrum of the spiral resonator below 1GHz at different temperatures. Ten modes, each with potential to create negative effective permeability, are shown."} \label{Resonance}
\end{figure}

Because of the capacitance between adjacent lines, and the magnetic and kinetic inductances along the spiral, the spiral resonator is a compact self-resonant device. As such, it exhibits a series of resonances corresponding to different modes of RF current standing waves, and these have been recently imaged by Laser Scanning Microscopy (LSM) in Nb spiral resonators\cite{Zhuravel}. Figure \ref{Resonance} depicts the transmission spectrum of our YBCO spiral resonator at several temperatures. Two channels contribute to the transmission spectrum: the direct coupling between the two loop antennas and the indirect coupling through the mutual inductances to the spiral resonator. Therefore, each resonance comprises a window of enhanced transmission followed by a window of suppressed transmission due to the constructive and destructive interference between these two channels, respectively\cite{Ghamsari}. The resonances are approximately located as $(2n-1)\Delta f$ in the frequency domain, where $n$=1,2,3,... is the mode order and $\Delta f\approx50MHz$ is the separation between the modes. Thus, the spiral resonator is a multi-band metamaterial atom where each individual mode offers a narrow-band window of effective negative-permeability.

The fundamental mode, at $53.4MHz$, possesses the lowest quality factor, $Q=1040$ at 10K, among the modes due to its large dipole magnetic moment and strong coupling to the loop antennas. Much higher quality factors are achievable at higher order modes, e.g $Q\approx30000$ for the sixth mode, where the net magnetic moment of the resonant mode is smaller.

\begin{figure}
\centering{\includegraphics[width=3.5in]{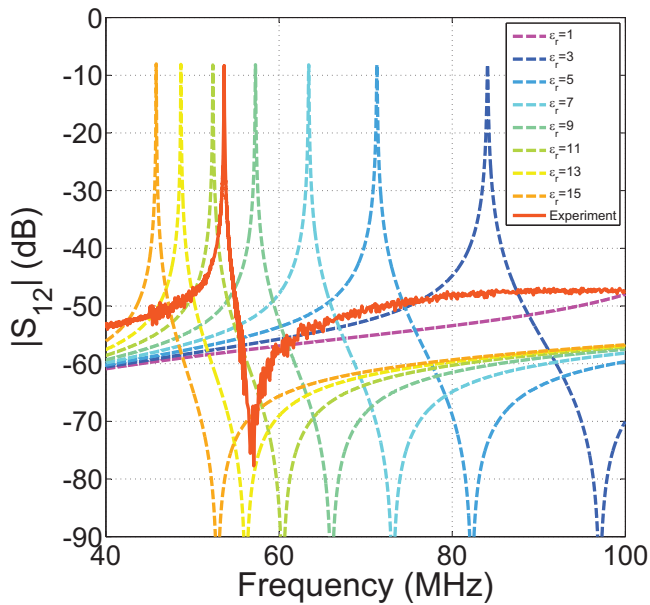}
} \caption{The effect of substrate dielectric constant on the resonance frequency of the fundamental mode. The solid line shows the experimental resonance of the YBCO spiral on sapphire substrate and the dashed lines are the results of HFSS simulation of a perfect metal spiral of the same geometry with varying dielectric constant in the substrate.} \label{Substrate}
\end{figure}

In order to optimize the device, the effect of the substrate on the performance of the spiral resonator is examined by measuring similar devices on different substrates, namely MgO, Al$_2$O$_3$, and LaAlO$_3$.
The immediate effect of the substrate on the resonance spectrum is to change the resonance frequencies due to the specific dielectric constant of the substrate, which modifies the effective capacitance between the spiral turns.
This point is illustrated by HFSS simulation in Figure \ref{Substrate} and is evident from the experiments as summarized in Table \ref{fQ}.

Nonetheless, the substrate affects the resonances in a more profound way through its influence on the morphology of the superconducting film. For example, the LaAlO$_3$ substrate contains strongly twinned domains; therefore, superconducting film grown on it also suffers from twinning. As has been shown earlier by laser scanning microscopy\cite{Kurter2,Zhuravel2}, the border of adjacent twinned domains is vulnerable to enhanced photoresponse, creation of hot spots and is a host for dissipation and nonlinearities. Hence, one expects devices on LaAlO$_3$ substrates to exhibit a lower quality factor, Q, than similar devices on MgO or Al$_2$O$_3$. Table \ref{fQ} clearly shows this fact, indicating significant improvement on the device performance when un-twinned substrates are used.

In addition, patterning deficiencies and geometrical defects reduce the inductance of the spiral, and tend to shift the resonance to higher frequencies. This effect could compensate for, or even dominate over, the increased capacitance gained by using a high-$\epsilon_r$ substrate, like LaAlO$_3$, such that the net effect will be pushing the resonances to higher frequencies and widening the separation between adjacent modes. Therefore, refining the patterning technique for each individual substrate is very important in order to get a high quality meta-atom.

We have demonstrated fabrication and measurement of high-temperature superconducting spiral resonators for metamaterial applications at RF frequencies.
In particular, the devices enjoy a deep sub-wavelength feature size, which enables them to be cast into arrays by standard micro-fabrication techniques.
Moreover, YBCO metamaterial atoms have the advantage of providing a wide range of operating temperature, which facilitates tunability of the device.
It was shown that the device can be utilized as a multi-band metamaterial atom due to its regular spectrum of resonances.
The effect of substrate was studied experimentally by implementing the same device on MgO, Al$_2$O$_3$, and LaAlO$_3$ substrates.
We noticed that devices on un-twinned substrates, like sapphire, systematically exhibit higher quality factors for all the resonances.

This work is supported by the DOE grant number DESC0004950 and ONR/UMD/AppEl, Task D10, through grant number N000140911190.
We acknowledge early assistance from Cihan Kurter.

\end{document}